\documentclass[11pt]{article}

\begin{document}

\title{V.M. Red'kov\footnote{redkov@dragon.bas-net.by} \\
Spherical waves for Dirac--K\"{a}hler and Dirac particles,\\
formal relations between boson and fermion
solutions\footnote{Translated version of a paper: VINITI 16.08.89,
no 5315 - B89, Minsk,  1989; Chapter 2 in:
 V.M. Red'kov. Tetrad formalism, spherical symmetry and Schr\"{o}dinger basis.
Publishing House "Belarusian Science", Minsk, 339 pages (2011).}\\
{\small B.I. Stepanov Institute of physics, National Academy of
Sciences of Belarus} }

\date{}
\maketitle

\begin{quotation}

Tetrad based equation for Dirac--K\"{a}hler particle is solved in spherical coordinates in the flat Minkocski space-time.
Spherical solutions  of boson type ($J =0,1,2,...$) are constructed.
After performing a special transformation over spherical boson
solutions of the Dirac--K\"{a}hler equation, $4 \times 4$-matrices $U(x) \Longrightarrow V(x)$,  simple linear expansions of the four rows of new representative
of the Dirac--K\"{a}hler field $V(x)$ in terms of spherical fermion solutions $\Psi_{i}(x)$  of the four ordinary Dirac equation
have been derived. However, this fact cannot be interpreted as the  possibility not to  distinguish between
the Dirac-K\"{a}hler field and the system four Dirac fermions.
The main formal argument  is that the special transformation $(I \otimes S(x))$ involved  does not belong
to the group of tetrad local gauge transformation for Dirac-K\"{a}hler field, 2-rank bispinor under the Lorentz group.
Therefore, the linear expansions between boson and fermion functions are not gauge invariant under the  group of local tetrad rotations.

\end{quotation}

\section{Spherical solution of the Dirac--K\"{a}hler}

The Dirac--K\"{a}hler field (other terms are Ivanenko--Landau field or vector field of general type)
was investigated by many authors -- see bibliography in \cite{paper-1, Book-2}. In the context of the most
intriguing  question -- what does describe  this field,  a boson or
a composite fermion with internal degree of  freedom,
the goal of  the paper is to construct spherical solutions of the
Dirac--K\"{a}hler field both of boson and fermion type and then to describe
relations between them. The problem is solved  in the flat Minkowski space.
Additionally  we specify the case of a curved space time background (3-space of constant positive curvature) where
any  fermion solutions cannot be constructed.

The Dirac--K\"{a}hler,  written in a diagonal spherical tetrad of the flat Minkowski space-time,
has the form
$$
\left  [ \; i \gamma ^{0}\; \partial _{t} \; + \;
 i\; \;( \gamma ^{3} \; \partial _{r} \; +  \; { \gamma ^{1}J^{31} \; + \;
\gamma ^{2}\; J^{32} \over r}\; ) \; + \; {1 \over r}\;
 \Sigma _{\theta ,\phi }\; - \; m\;  \right  ]\;  U(x) = 0 \; ,
\eqno(1.1a)
$$
$$
\Sigma _{\theta ,\phi }  =  \; i \gamma ^{1}\; \partial _{\theta }
\; +  \; \gamma ^{2}\; {i\partial _{\phi }\; +\; iJ^{12} \; \cos
\theta \over  \sin  \theta} \;  , \qquad  J^{12} = ( \sigma ^{12}
\otimes  I \; + \;  I \otimes  \sigma ^{12} ) \; . \eqno(1.1b)
$$

\noindent By diagonalizing operators  ${\bf J}^{2}, J_{3}$ of the total angular momentum
(first constructing solutions of boson type)
$$
J_{1}  =  l_{1} \;+\; {i J^{12} \cos \phi \over  \sin \theta } \;
, \;\; J_{2} =   l_{2} \;+\; {i J^{12} \sin \phi \over  \sin
\theta }  \; , \;\; J_{3} = l_{3} \; , \eqno(1.2a)
$$

\noindent for the wave function we obtain substitution (details of the relevant general formalism see in \cite{Book-2})
$$
U_{\epsilon JM}(t,r,\theta ,\phi ) = { e^{-i\epsilon t} \over  r}
\; \left | \begin{array}{llll}
f_{11} \; D_{-1} & f_{12} \; D_{0} & f_{13} \; D_{-1} & f_{14} \; D_{0} \\
f_{21} \; D_{0}  & f_{22} \; D_{+1}& f_{23} \; D_{0}  & f_{24} \; D_{+1} \\
f_{31} \; D_{-1} & f_{32} \; D_{0} & f_{33} \; D_{-1} & f_{34} \; D_{0} \\
f_{41} \; D_{0}  & f_{42} \; D_{+1}& f_{43} \; D_{0}  & f_{44} \;
D_{+1}
\end{array} \right |     ,
\eqno(1.2b)
$$

\noindent  $f_{ab} = f_{ab}(r)\; ,\; D_{\sigma } =
D^{J}_{-M,\sigma } (\phi ,\theta ,0)$,  a quantum number  $J$
takes in the values  $0, \; 1, \; 2, \ldots$ When calculating the action of angular operator,
$\Sigma _{\theta ,\phi }\; U_{\epsilon JM}$, we need to employ the known formulas
\cite{Varshalovich-Moskalev-Hersonskiy-1975}
$$
\partial _{\theta }\; D_{-1} = {1 \over 2}
(b \; D_{-2} - a\; D_{0})\; , \qquad  [(-M + \cos \theta ) / \sin
\theta ] \;
 D_{-1} = {1 \over 2} (-b \; D_{-2} -a\; D_{0}) \; ,
$$
$$
\partial _{\theta }\; D_{+1} = {1 \over 2} (a\; D_{0} - b\; D_{+2}) \; ,\qquad
[(-M - \cos \theta ) / \sin \theta ] \; D_{+1} = {1 \over 2}
 (-a \; D_{0} - b \; D_{+2})\; ,
$$
$$
\partial _{\theta }\; D_{0} = {1 \over 2} (a\; D_{-1} - a \; D_{+1}) \; ,\;\qquad
 [-M / \sin \theta ] \; D_{0}  = {1 \over 2} ( -a\; D_{-1} - a\; D_{+1})\; , \;
$$
$$
 a = \sqrt{J(J+1)}\; , \qquad  b = \sqrt{(J-1)(J+1) } \; .
\eqno(1.3a)
$$

\noindent So, for  $\Sigma _{\theta ,\phi }\; U_{\epsilon JM}$ we get
$$
\Sigma _{\theta ,\phi } \; U  = i \sqrt{J(J+1)} \left |
\begin{array}{rrrr} - f_{41}\;  D_{-1}  & - f_{42} \; D_{0}  &
 - f_{43} \; D_{-1} & - f_{44} \; D_{0}  \\
f_{31}  \;  D_{0}   &   f_{32} \; D_{+1}  &
f_{33}  \;  D_{0}   &   f_{34} \; D_{+1} \\
f_{21}  \;  D_{-1}  &   f_{22} \; D_{0}   &
f_{23}  \;  D_{-1}  &   f_{24} \; D_{0}  \\
- f_{11}\;  D_{0}   &  -f_{12} \; D_{+1}  & - f_{13}\;  D_{0}   &
-f_{14} \; D_{+1}
\end{array} \right |       \; .
\eqno(1.3b)
$$

To simplify the problem, on the functions  $U_{\epsilon JM}$  let us diagonalyze additionally
an operator of $P$-reflection for the Dirac--K\"{a}hler field; in the basis of spherical tetrad it
has the form
$$
\hat{\Pi}_{sph} =  \left | \begin{array}{cccc}
0 &  0 &  0 &  -1  \\    0 &  0 & -1 &   0 \\
0 & -1 &  0 &   0  \\   -1 &  0 &  0 &   0
\end{array} \right |     \otimes     \left | \begin{array}{cccc}
0 &  0 &  0 &  -1  \\    0 &  0 & -1 &   0 \\
0 & -1 &  0 &   0  \\   -1 &  0 &  0 &   0    \end{array} \right |
\otimes \hat{P} \; \; . \eqno(1.4a)
$$

\noindent From eigenvalue equation  $ \hat{\Pi}_{sph} U_{\epsilon JM} = \Pi \; U_{\epsilon
JM}$, we derive the following restrictions

$$
f_{31}  = \pm  f_{24} \;\;  , \;\; f_{32} = \pm  f_{23} \;\; ,
\;\; f_{33}  = \pm  f_{22} \;\;  , \;\; f_{34} = \pm  f_{21} \;\;
,
$$
$$
f_{41}  = \pm  f_{14} \;\; , \;\; f_{42} = \pm  f_{13} \;\; , \;\;
f_{43}  = \pm  f_{12} \;\; ,\;\; f_{44} = \pm  f_{11} \;\; .
\eqno(1.4b)
$$

\noindent  upper  sign  concerns the value $\Pi = (-1)^{J+1}$,
lower concerns the values  $\Pi = (-1)^{J}$. Below, to distinguish between two cases  with different parity,
we will refer  $\Pi =
(-1)^{J+1}$  with the symbol  $\Delta = - 1$, and  $\Pi = (-1)^{J}$ with  $\Delta  = + 1$.

Correspondingly, the substitution will read
$$
U_{\epsilon JM\Delta } (t,r,\theta ,\phi ) = {e^{-i\epsilon t}
\over r} \left | \begin{array}{rrrr}
f_{11} \; D_{-1} & f_{12} \; D_{\;\;0}  & f_{13} \; D_{-1} & f_{14} \; D_{\;\;0}   \\
f_{21} \; D_{\;\;0}  & f_{22} \; D_{+1} & f _{23}\; D_{\;\;0}  & f_{24} \;  D_{+1} \\
\Delta  f_{24} \; D_{-1}  & \Delta  f_{23} \; D_{\;\;0}  &
\Delta  f_{22} \; D_{-1}  & \Delta  f_{21} \; D_{\;\;0}  \\
\Delta  f_{14} \; D_{\;\;0}   & \Delta  f_{13} \; D_{+1} & \Delta
f_{12} \; D_{\;\;0}   & \Delta  f_{11} \; D_{+1} \noindent
\end{array} \right |     \; . \eqno(1.5a)
$$

The system of radial equations at $\Delta = + 1$  is
$$
\epsilon  f_{24}  - i {d  \over dr} f_{24} - i {a \over r} f_{14}
- m f_{11}  = 0  \; ,
$$
$$
\epsilon  f_{23} - i{d  \over dr} f_{23} + i{1 \over r}f_{14} - i
{a \over r} f_{13} - m f_{12} = 0 \; ,
$$
$$
\epsilon  f_{22} - i {d \over dr} f_{22} - i {a \over r} f_{12} -
m f_{13}  = 0 \; ,
$$
$$
\epsilon  f_{21} - i { d \over dr}f_{21} + i {1\over r} f_{12}  -
i {a \over r} f_{11}  - m f_{14}  = 0 \; ,
$$
$$
\epsilon  f_{14} + i {d \over dr} f_{14} + i {1 \over r} f_{23} +
i {a\over r} f_{24}  - m f_{21}  = 0 \; ,
$$
$$
\epsilon  f_{13} + i {d \over dr} f_{13}  + i {a \over r} f_{23}
- m f_{22}  = 0  \; ,
$$
$$
\epsilon  f_{12} + i {d  \over dr} f_{12} + i {1 \over r} f_{21} +
i {a \over r} f_{22}  - m f_{23} = 0 \; ,
$$
$$
\epsilon  f_{11} + i {d \over dr} f_{11} + i {1 \over r} f_{21} +
i {a \over r} f_{21}  - m f_{24}  = 0 \; . \eqno(1.5b)
$$

\noindent Having changed  $m$  into    $-m$,  we produce analogous equations at
  $\Delta  = - 1$.

  Now let us translate equations to the following new combinations

  $$
A = (f_{11} + f_{22}) / \sqrt{2} \; , \qquad B = (f_{11} - f_{22})
/ i\sqrt{2}   \;  ,
$$
$$
C = (f_{12} + f_{21}) / \sqrt{2} \;, \qquad D = (f_{11} - f_{22})
/ i\sqrt{2} \; ,
$$
$$
K = (f_{13} + f_{24}) / \sqrt{2} \; , \qquad L = (f_{13} - f_{24})
/ i\sqrt{2} \;  ,
$$
$$
M = (f_{14} + f_{23}) / \sqrt{2}  \; , \qquad N = (f_{14} +
f_{23}) / i\sqrt{2} \; . \eqno(1.6a)
$$

\noindent As result, we  obtain equations without imaginary $i$

$$
\epsilon  K  - {dL \over dr} + {a \over r} N  - m A   = 0 \; ,
$$
$$
\epsilon  L  + {dK \over dr} + {a \over r} N  + m B  = 0 \; ,
$$
$$
\epsilon  A - {dB \over dr} + {a \over r} D  - m K  = 0 \; ,
$$
$$
\epsilon  B + {dA \over dr} + {a \over r} C + m L  = 0  \; ,
$$
$$
\epsilon  M - {dN \over dr} + {1\over r} N + {a \over r} L - m C =
0 \; ,
$$
$$
\epsilon  N + {dM \over dr} + {1 \over r} M + {a \over r} K + m D
= 0 \; ,
$$
$$
\epsilon  C - {dD \over dr} + {1 \over r} D + {a \over r} B - m M
= 0 \; ,
$$
$$
\epsilon  D + {dC \over dr} + {1 \over r} C + {a \over r} A + m N
= 0 \; . \eqno{1.6b}
$$

Note. that eqs.     $(1.6b)$ permit the following linear constraints
$$
A = \lambda \;  K \; , \;\; B = \lambda \; L \; , \;\; C = \lambda
\;  M \; , \;\; D = \lambda \;  N \; , \eqno(1.7a)
$$

\noindent where  $\lambda  = \pm 1$. In particular, at $\lambda = + 1$
we get a system of four  equations
$$
{dK \over dr} + {a \over r} M + (\epsilon  + m) L = 0 \; ,
$$
$$
{dL \over dr} - {a \over r} N - (\epsilon  - m) K = 0 \;  ,
$$
$$
({d \over dr} + {1 \over r} ) M  + {a \over r} K + (\epsilon  + m)
N = 0 \; ,
$$
$$
 ({d \over dr} - {1 \over r} ) N  - {a \over r} L
- (\epsilon  - m) M = 0 \; . \eqno(1.7b)
$$

\noindent By formal changing  $m$ into  $-m$,  we  will obtain equations for the case   $\lambda = -1$.

Taking these restrictions into account, the substitution for solution $U^{\lambda }_{\epsilon JM\Delta }$
 can be written in a simpler form
$$
U^{\lambda }_{\epsilon JM\Delta } (t,r,\theta ,\phi ) =
{e^{-i\epsilon t} \over  r \sqrt{2}}   \times
$$
$$
\left | \begin{array}{rrrr} \lambda\; (K+iL)\; D_{-1} & \lambda\;
(M+iN)\; D_{\;\;0} &
(K+iL)\; D_{-1} & (M+iN)\; D_{\;\;0} \\[2mm]
\lambda\; (M-iN)\; D_{\;\;0} & \lambda\; (K-iL)\; D_{+1} &
(M-iN)\; D_{\;\;0} & (K-iL)\; D_{+1} \\[2mm]
\Delta\; (K-iL)\; D_{-1}  &  \Delta\; (M-iN)\; D_{\;\;0}  &
\Delta\; \lambda\; (K-iL)\; D_{-1} &  \Delta\; \lambda\; (M-iN)\; D_{\;\;0} \\[2mm]
\Delta\;  (M+iN)\; D_{\;\;0}  &  \Delta \; (K+iL)\; D_{+1}  &
\Delta \; \lambda\; (M + iN)\; D_{\;\;0}  & \Delta\;  \lambda\;
(K+iL)\; D_{+1}
\end{array} \right |      \; .
$$
$$
\eqno(1.7c)
$$

Equations  $(1.7b)$ can be solved with the use of two different substitutions

$$
I. \;\;  \sqrt{J+1} \; K(r) = f(r)\;\; , \;\; \sqrt{J+1} \;\; L(r)
= g(r) \; \; ,
$$
$$
\sqrt{J}\; M(r) = f(r) \;\; , \;\; \sqrt{J} \; N(r) = g(r) \; ;
\eqno(1.8a)
$$

$$
II. \;\;   \sqrt{J}\; K(r) = f(r)\;\; , \;\; \sqrt{J}\; L(r) =
g(r) \; \;,
$$
$$
\sqrt{J+1}\; M(r) = - f(r)\;\; , \; \; \sqrt{J+1} \; N(r) = - g(r)
\;  . \eqno(1.8b)
$$

\noindent In the case  $(1.8a)$, we get

$$
I. \qquad ({d \over dr} + {J+1 \over r}) f + (\epsilon + m) g  = 0
\; ,
$$
$$
\qquad ({d \over dr} - {J+1 \over r}) g -
(\epsilon - m) f = 0 \; . \eqno(1.9a)
$$

\noindent and similarly for  $(1.8b)$  we obtain

$$
II. \qquad ( {d \over dr}  - {J \over r} ) f + (\epsilon + m) g  = 0
\; ,
$$
$$
\qquad ({d \over dr} + {J \over r} ) g  -
(\epsilon - m) f = 0 \;  . \eqno(1.9b)
$$

\noindent Remember that eqs.   $(1.9a,b)$ refer to the case  $\Delta = + 1$  и $\lambda  = + 1$.

Thus, at fixed quantum numbers  $(\epsilon ,\; J,\; M,\;\Delta )$, there exist  four types
of solutions: due to two number for $\lambda  = \pm
1$ and due to existence of two substitutions  $I$  and  $II$  (see  $ (1.8a,b)$).

 Solutions of the type    $I$ are described by
 $$
U^{I. \lambda } _{\epsilon JM\Delta }(x) = { e^{i \epsilon t}
\over r }     \times
$$
$$
\left | \begin{array}{cccc} \lambda \; D_{-1} / \sqrt{J+1} &
\lambda \; D_{\;\;0} / \sqrt{J}   &
D_{-1} / \sqrt{J+1}  &   D_{\;\;0} / \sqrt{J}   \\
\lambda  \; D_{\;\;0}  / \sqrt{J}    & \lambda  \; D_{+1} /
\sqrt{J+1}  &
D_{\;\;0}/\sqrt{J}    &    D_{+1} /  \sqrt{J+1}   \\
D_{-1} / \sqrt{J+1} & D_{\;\;0} /  \sqrt{J}  & \lambda \; D_{-1} /
\sqrt{J+1} &
\lambda \; D_{\;\;0} / \sqrt{J}  \\
D_{\;\;0} / \sqrt{J}  &  D_{+1} / \sqrt{J+1}  & \lambda\;
D_{\;\;0} /\sqrt{J}   &   \lambda \; D_{+1} / \sqrt{J+1}
\end{array} \right |       \left. \begin{array}{l}
\leftarrow \;\;\; (f+ig) / \sqrt{2}  \\
\leftarrow \;\;\; (f-ig) / \sqrt{2}  \\
\leftarrow \Delta (f-ig) / \sqrt{2}  \\
\leftarrow \Delta (f+ig) / \sqrt{2}
\end{array} \right.        \; ;
$$
$$
\eqno(1.10a)
$$

\noindent where all elements of each line should by multiplied by
 a function from the left; at  $\Delta = + 1 , \lambda  = + 1$,
the functions  $f$ and  $g$ obey eqs. $(1.9a)$, whereas for three remaining cases
we should perform in  $(1.9a)$ formal changes in accordance with the rules

$$
(\; \Delta  = - 1, \; \lambda  = + 1\; ) \qquad   m \; \rightarrow
\;  - m  \; ;
$$
$$
(\; \Delta  = + 1, \; \lambda  = - 1\; )  \qquad   m \;
\rightarrow  - m     \; ;
$$
$$
(\; \Delta  = - 1, \; \lambda  = - 1\; )  \qquad   m \;
\rightarrow  + m \;  . \eqno(1.10b)
$$

\noindent Analogously, for solutions of the second type
we have
$$
U^{II.\; \lambda }_{\epsilon JM\Delta } (x)  = { e^{-i\epsilon t}
\over r } \times
$$
$$
\left | \begin{array}{cccc} -\lambda \; D_{-1} / \sqrt{J}   &
 \lambda \; D_{\;\;0} / \sqrt{J+1}  &
D_{-1}  / \sqrt{J}             &
-D_{\;\;0} /  \sqrt{J+1} \\
\lambda \; D_{\;\;0} / \sqrt{J+1}  & -\lambda\; D_{+1} / \sqrt{J}
& - D_{\;\;0} / \sqrt{J+1}  &
D_{+1} /  \sqrt{J}  \\
D_{-1} /  \sqrt{J}  & - D_{\;\;0} / \sqrt{J+1}  & - \lambda \;
D_{-1} / \sqrt{J}  &
\lambda \; D_{\;\;0} / \sqrt{J+1}    \\
- D_{\;\;0} / \sqrt{J+1}  & D_{+1} /  \sqrt{J}   & \lambda\; D
_{\;\;0} / \sqrt{J+1}  & -\lambda \; D_{+1} / \sqrt{J}
\end{array} \right |     \left. \begin{array}{l}
\leftarrow \;\;\;(f+ig) /  \sqrt{2}   \\
\leftarrow \;\;\;(f-ig) /  \sqrt{2}   \\
\leftarrow \Delta(f-ig) /  \sqrt{2}   \\
\leftarrow \Delta(f+ig) /  \sqrt{2}
\end{array} \right.                 \; .
$$
$$
\eqno(1.10c)
$$

\noindent At $\Delta  = + 1 ,\; \lambda  = +  1$, the functions  $f$ and $g$
 obey eqs. $(1.9b)$; in three remaining cases one should use the rules
  $(1.10b)$.

The case of minimal value  $J = 0$  needs a separate consideration.
Indeed, initial substitution for the wave function $U_{\epsilon 00}(x)$   turns to be independent on
angular variables
$$
U_{\epsilon 00}(t,r) = {e^{-i\epsilon t} \over r} \left |
\begin{array}{cccc}
0 & f_{12} & 0 & f_{14} \\ f_{21} & 0 & f_{23} & 0 \\
0 & f_{32} & 0 & f_{34} \\ f_{41} & 0 & f_{43} & 0
\end{array} \right |       \; .
\eqno(1.11)
$$

\noindent The operator of spacial inversion, being only a matrix operation,
permits to separate functions (1.11) in two classes -- eigenvalue equation
 $\hat{\Pi}\; U_{\epsilon 00} = \Pi \; U_{\epsilon 00}$  gives:

\vspace{3mm}

  $ \Pi = + 1\;  ( \Delta  = + 1  ) ,$
$$
f_{32} = + f_{23} \; ,\;\; f_{34} = + f_{21} \; , \;\; f_{41} = +
f_{14} \; , \;\; f_{43} = + f_{12}     \;\; ; \eqno(1.12a)
$$

 $\Pi = - 1 ( \Delta  = - 1  ),$
$$
 f_{32} = - f_{23} \; ,
\;\; f_{34} = - f_{21} \; ,  \;\; f_{41} = - f_{14} \; , \;\;
f_{43} = - f_{12} \;\;  . \eqno(1.12b)
$$

\noindent Allowing for relation  $\Sigma _{\theta ,\phi } \; U_{\epsilon
00} =  0$,  we derive the radial system
(for  states with  $J09$ , functions  $A
,\; B ,\; K ,\; L$  in $(1.6a)$)  vanish identically)
$$
\epsilon\;  M  - {dN \over dr} + {N \over r} - m\; C  = 0 \;\; ,
$$
$$
 \epsilon\;  N  + {dM \over dr} + {M \over r} + m\; D  = 0  \;
,
$$
$$
\epsilon\;  C  - {dD \over dr} + {D \over r} - m \; M  = 0  \; ,
$$
$$
 \epsilon\;  D + {dC \over dr}  + {C \over r} - m \; N  =  0
\;   . \eqno(1.12c)
$$

\noindent To obtain equations when  $\Delta =
- 1$, one should change  $m$  into   $-m$.

The system  (1.12) can be simplified by two substitutions:

\vspace{3mm} $ C= + M \; , \; D = + N \;  ( \lambda  = + 1 ) $
$$ ( {d \over
dr} + {1 \over r} ) M + ( \epsilon  + m ) N  = 0 \; ,
$$
$$
 ( {d \over dr} - {1 \over r} ) N - ( \epsilon  - m ) M = 0  \;  ;
\eqno(1.13a)
$$

 $ C = - M , \; D = - N\;  (\lambda = - 1)  $
$$
( {d\over dr}  + {1 \over r} ) M + ( \epsilon  - m ) N = 0 \; ,
$$
$$
( {d \over dr} - {1 \over r} ) N - ( \epsilon  + m ) M = 0
\; . \eqno(1.13b)
$$

Thus, at  $J = 0$ and a fixed parity, there exist two different solutions   (doubling by  $\lambda=\pm 1$):
$$
U^{\lambda } _{\epsilon 00\Delta }(t,r) = {e^{-i\epsilon t} \over
r} \left | \begin{array}{cccc}
0  & \lambda  & 0 & 1 \\
\lambda  & 0 & 1 & 0 \\
0 & 1 & 0 & \lambda  \\
1 & 0 & \lambda  & 0     \end{array} \right | \left.
\begin{array}{l}
\leftarrow \;\;\; (M + i N)/\sqrt{2}    \\
\leftarrow \;\;\; (M - i N)/\sqrt{2}    \\
\leftarrow \Delta (M - i N)/\sqrt{2} \\
\leftarrow \Delta (M + i N)/\sqrt{2}      \end{array} \right. \; .
\eqno(1.14)
$$

\section{ On relations between boson and fermion solutions}

Now let us relate the above spherical solutions of boson type with spherical
solutions of  the ordinary Dirac equation \cite{Book-2}
$$
\Psi _{\epsilon jm\delta }(x)  = { e^{-i\epsilon t} \over r }
\left | \begin{array}{c} D_{-1/2} \\[2mm]   D_{+1/2} \\[2mm]   D_{-1/2} \\[2mm]
D_{+1/2}    \end{array} \right | \left. \begin{array}{l}
\leftarrow \;\; (F + i G)/\sqrt{2}   \\[2mm]
\leftarrow \;\; (F - i G)/\sqrt{2}   \\[2mm]
\leftarrow \delta  (F - i G)/\sqrt{2} \\[2mm]
\leftarrow \delta  (F + i G)/\sqrt{2}  \end{array} \right.\; ,
\eqno(2.1a)
$$

\noindent where  $\delta = + 1$ refers to the parity $P =
(-1)^{j+1}$, and  $\delta = - 1$ refers to the parity $P =
(-1)^{j}$. Radial equations for  $F$ and  $G$ at $ \delta = + 1  $
are
$$
\left ( {d \over dr} +  {j + 1/2 \over r}  \right )  F + (
\epsilon  + m )\; G = 0 \; ,
$$
$$
\left ( {d \over dr}  - {j + 1/2 \over r} \right ) G  - ( \epsilon
- m )\; F = 0\; ; \eqno(2.1b)
$$

\noindent changing the sign at  $m$ in  $(2.1b)$ we obtain equation for states with  different parity
 (the case $\delta = - 1$).

In order to connect explicitly the above boson solutions of the Dirac--K\"{a}hler field
with  spherical solutions of the (four) Dirac equations, one must
perform over the matrix   $U(x)$ a special  transformation $U(x) \;
\rightarrow \; V(x)$, choosing it so that in a new representation the Dirac--K\"{a}hler equation is splitted into four
separated Dirac-Like equation;  then there arises possibility
 to decompose four rows of the $4 \times 4$-matrix  $V(x)$, related with  the Dirac-K\"{a}hler  equation,
 in terms of  solutions of four  Dirac equation.

%  Тем  самым  в  линейном пространстве $
%\{ \; ( \; \Psi ^{1} , \; \Psi ^{2} ,\; \Psi ^{3} ,
%   \; \Psi ^{4}\; ) ; \; \Psi ^{i} - \;\; \mbox{решения} \;\; \mbox{уравнения}
%   \;\; \mbox{Дирака}
%\; \} $
% можно будет явно выделить подпространство, в
%определенном  смысле отвечающее частице Дирака--Кэлера. Нужное
%калибровочное преобразование  имеет вид
The transformation we need has the form
$$
V(x)  = ( I \otimes  S(x))\; U(x) \; , \qquad  S(x)  = \left |
\begin{array}{cc}  B(x) & 0 \\ 0 & B(x)
\end{array} \right |  \; ,
$$

$$
B(x) =   \left |  \begin{array}{cc}
\cos {\theta \over 2} \; e^{-i\phi /2} & \sin  {\theta \over 2} \; e^{-i\phi /2} \\[2mm]
-\sin  {\theta \over 2} \; e^{+i\phi /2} & \cos {\theta \over 2}\;
e^{+i\phi /2}
\end{array}  \right |     \; .
\eqno(2.2a)
$$

\noindent Spherical bispinor connection  $\Gamma _{\alpha }$

$$
\Gamma _{t} = 0 \; , \;\; \Gamma _{r} = 0 \; , \;\; \Gamma
_{\theta } = \sigma ^{12} \; ,  \;\; \Gamma _{\phi } =  \sin
\theta \; \sigma ^{32}  +  cos \theta \;\sigma ^{12}
$$

\noindent entering the Dirac--K\"{a}hler equation translated to $V(x)$-representation
$$
 \{ \; [\; i \gamma ^{\alpha }(x) \; ( \partial_{\alpha } \; + \;
\Gamma _{\alpha }(x))  \; - \; m \;  ] \; V(x)
$$
$$
 + i\; \gamma ^{\alpha }(x) \; V(x) \; [\; S(x) \; \Gamma
_{\alpha}(x) \; S^{-1}(x) \; + \; S(x) \; \partial _{\alpha}  \;
S^{-1}(x) \; ]\;  \}  = 0 \eqno(2.2b)
$$

\noindent will make to zero the following term
$$
 S(x)\; \Gamma _{\alpha }(x)\; S^{-1}(x) \; + \;
S(x)\; \partial_{\alpha }\;  S^{-1}(x) = 0
$$

\noindent and we obtain what needed
$$
\left [ \; i \gamma ^{\alpha }(x)\; (\partial _{\alpha }\; + \;
\Gamma _{\alpha }(x)) \; -  \; m \; \right ]\;  V(x) = 0\; .
\eqno(2.2c)
$$

 Thus, the task consists in the following:

 \vspace{5mm}

1)  first, we should translate the above spherical solutions in  $U$-form to corresponding $V$-form;

 2)  second,  we should  expand  four rows of the matrix $V$ in terms of Dirac spherical waves.

\vspace{5mm}

With the use of   (2.2a), the matrix  $ U_{\epsilon JM}$ in  (1.3b) will assumes the forn
 (we have written  $V_{ij}$  by   rows)

$$
(V_{i1}) =   \left | \begin{array}{c}
 f_{11} D_{-1} \cos {\theta \over 2} e^{-i\phi /2} \;  - \;
f_{12} D_{0}  \sin {\theta \over 2} e^{-i\phi /2}      \\[2mm]
 f_{21} D_{0}  \cos {\theta \over 2} e^{-i\phi /2} \;  - \;
f_{22} D_{+1} \sin {\theta \over 2} e^{-i\phi /2}     \\[2mm]
 f_{31} D_{-1} \cos {\theta \over 2} e^{-i\phi /2} \;  -  \;
f_{32} D_{0}  \sin {\theta \over 2} e^{-i\phi /2}     \\[2mm]
 f_{41} D_{0}  \cos {\theta \over 2} e^{-i\phi /2} \;  - \;
f_{42} D_{+1} \sin {\theta \over 2} e^{-i\phi /2}
\end{array} \right | \; ,
$$

$$
(V_{i2})  =    \left | \begin{array}{c}
 f_{11} D_{-1} \sin {\theta \over 2} e^{+i\phi /2} \; + \;
f_{12} D_{0}  \cos {\theta \over 2} e^{+i\phi /2}      \\[2mm]
 f_{21} D_{0}  \sin {\theta \over 2} e^{+i\phi /2} \; +  \;
  f_{22} D_{+1} \cos {\theta \over 2} e^{+i\phi /2}          \\[2mm]
 f_{31} D_{-1} \sin {\theta \over 2} e^{+i\phi /2} \; + \;
f_{32} D_{0}  \cos {\theta \over 2} e^{+i\phi /2}     \\[2mm]
f_{41} D_{0}  \sin {\theta \over 2} e^{+i\phi /2} \; + \; f_{42}
D_{+1} \cos {\theta \over 2} e^{+i\phi /2}
\end{array} \right | \; ,
$$

$$
(V_{i3}) =     \left | \begin{array}{c}
 f_{13} D_{-1} \cos {\theta \over 2} e^{-i\phi /2}  \; - \;
f_{14} D_{0}  \sin {\theta \over 2} e^{-i\phi /2}  \\[2mm]
 f_{23} D_{0}  \cos {\theta \over 2} e^{-i\phi /2}  \; -
  f_{24} D_{+1} \sin {\theta \over 2} e^{-i\phi /2}    \\[2mm]
 f_{33} D_{-1} \cos {\theta \over 2} e^{-i\phi /2}   \; - \;
  f_{34} D_{0}  \sin {\theta \over 2} e^{-i\phi /2}    \\[2mm]
 f_{43} D_{0}  \cos {\theta \over 2} e^{-i\phi /2}    \; -
  f_{44} D_{+1} \sin {\theta \over 2} e^{-i\phi /2}
\end{array} \right | \; ,
$$

$$
(V_{i4}) =    \left | \begin{array}{c}
 f_{13} D_{-1} \sin {\theta \over 2} e^{+i\phi /2} \; + \;
  f_{14} D_{0}  \cos {\theta \over 2} e^{+i\phi /2}   \\[2mm]
 f_{23} D_{0}  \sin {\theta \over 2} e^{+i\phi /2} \; + \;
  f_{24} D_{+1} \cos {\theta \over 2} e^{+i\phi /2}   \\[2mm]
 f_{33} D_{-1} \sin {\theta \over 2} e^{+i\phi /2} \; + \;
  f_{34} D_{0}  \cos {\theta \over 2} e^{+i\phi /2}   \\[2mm]
 f_{43} D_{0}  \sin {\theta \over 2} e^{+i\phi /2} \; + \;
  f_{44} D_{+1} \cos {\theta \over 2} e^{+i\phi /2}
\end{array} \right | \; .
$$

We  will apply 8 formulas relating  $D$-functions of integer and half-integer $j$
\cite{1975-Varshalovich-Moskalev-Hersonskiy};  two of them are written down below
$$
\cos {\theta \over 2} e^{i\phi /2}\; D^{J}_{-M,0} = \sqrt{{J(J-M)
\over 2J+1}}\; D^{J-1/2} _{-M-1/2,-1/2} \; + \;
\sqrt{{(J+1)(J+M+1) \over 2J+1}}\;  D^{J+1/2}_{-M-1/2,-1/2} \;  ,
$$

$$
\cos {\theta \over 2} e^{i\phi /2} \; D^{J}_{-M,+1} =
\sqrt{{(J+1)(J-M) \over 2J+1}}\; D^{J-1/2} _{-M-1/2,+1/2}\; + \;
\sqrt{{J(J+M+1) \over 2J+1}} \; D^{J+1/2}_{-M-1/2,+1/2} \; .
$$

With the help of those relations, for  $(V_{ij})$ we obtain  (the factor  $e^{-i\epsilon t}/r$ is omitted)

$$
V_{\epsilon jm} =  \; V^{(J-1/2)} _{\epsilon JM} \;+\;
    V^{(J+1/2)} _{\epsilon JM} \;   ,
\eqno(2.3)
$$

\noindent where
$$
(V^{(J-1/2)}_{i1}) = \sqrt{{J+M \over 2J+1}} \left |
\begin{array}{c}
( \sqrt{J+1} f_{11} \; - \; \sqrt{J}   f_{12})\;  D^{J-1/2}_{-M+1/2,-1/2}  \\
( \sqrt{J}   f_{21} \; - \; \sqrt{J+1} f_{22})\;  D^{J-1/2}_{-M+1/2,+1/2}  \\
( \sqrt{J+1} f_{31} \; - \; \sqrt{J}   f_{32})\;  D^{J-1/2}_{-M+1/2,-1/2}  \\
( \sqrt{J}   f_{41} \; - \; \sqrt{J+1} f_{42})\;
D^{J-1/2}_{-M+1/2,+1/2}
\end{array} \right |    \; ,
$$
$$
(V^{(J-1/2)}_{i2}) = \sqrt{{J-M \over 2J+1}} \left |
\begin{array}{c}
-(\sqrt{J+1} f_{11} \; - \; \sqrt{J}   f_{12} ) \; D^{J-1/2}_{-M-1/2,-1/2}   \\
-(\sqrt{J}   f_{11} \; - \; \sqrt{J+1} f_{22} ) \; D^{J-1/2}_{-M-1/2,+1/2}   \\
-(\sqrt{J+1} f_{21} \; - \; \sqrt{J}   f_{22} ) \; D^{J-1/2}_{-M-1/2,-1/2}   \\
-(\sqrt{J}   f_{41} \; - \; \sqrt{J+1} f_{42} ) \;
D^{J-1/2}_{-M-1/2,+1/2}
\end{array} \right | \; ,\;\;\;
$$
$$
(V^{(J-1/2)}_{i3} ) = \sqrt{{J+M \over 2J+1}} \left |
\begin{array}{c}
( \sqrt{J+1} f_{13} \; - \; \sqrt{J}   f_{14} ) \; D^{J-1/2}_{-M+1/2,-1/2}  \\
( \sqrt{J}   f_{23} \; - \; \sqrt{J+1} f_{24} ) \; D^{J-1/2}_{-M+1/2,+1/2}   \\
( \sqrt{J+1} f_{33} \; - \; \sqrt{J}   f_{34} ) \; D^{J-1/2}_{-M+1/2,-1/2}   \\
( \sqrt{J}   f_{43} \; - \; \sqrt{J+1} f_{44} ) \;
D^{J-1/2}_{-M+1/2,+1/2}
\end{array} \right |    \; ,
$$
$$
(V^{(J-1/2)}_{i4})  = \sqrt{{J-M) \over  2J+1}} \left |
\begin{array}{c}
- ( \sqrt{J+1} f_{13} \; - \; \sqrt{J}   f_{14} ) D^{J-1/2}_{-M-1/2,-1/2}  \\
- ( \sqrt{J}   f_{23} \; - \; \sqrt{J+1} f_{24} ) D^{J-1/2}_{-M-1/2,+1/2}  \\
- ( \sqrt{J+1} f_{33} \; - \; \sqrt{J}   f_{34} ) D^{J-1/2}_{-M-1/2,-1/2}  \\
- ( \sqrt{J}   f_{44} \; - \; \sqrt{J+1} f_{44} )
D^{J-1/2}_{-M-1/2,+1/2}
\end{array} \right |
$$

\noindent and
$$
(V^{(J+1/2)}_{i2} ) = \sqrt{{J-M+1 \over 2J+1}} \left |
\begin{array}{c}
( \sqrt{J}   f_{11} \; + \; \sqrt{J+1} f_{12} ) \; D_{-M+1/2,-1/2}^{J+1/2}  \\
( \sqrt{J+1} f_{21} \; + \; \sqrt{J}   f_{22} ) \; D_{-M+1/2,+1/2}^{J+1/2}  \\
( \sqrt{J}   f_{31} \; + \; \sqrt{J+1} f_{32} ) \; D_{-M+1/2,-1/2}^{J+1/2}  \\
( \sqrt{J+1} f_{41} \; + \; \sqrt{J}   f_{42} ) \;
D_{-M+1/2,+1/2}^{J+1/2}
\end{array} \right |           \; ,
$$
$$
( V^{(J+1/2)}_{i2} )     = \sqrt{{J+M+1 \over 2J+1}} \left |
\begin{array}{c}
( \sqrt{J}   f_{11} \; + \; \sqrt{J+1} f_{12} ) \; D^{J+1/2}_{-M-1/2,-1/2}  \\
( \sqrt{J+1} f_{11} \; + \; \sqrt{J}   f_{22} ) \; D^{J+1/2}_{-M-1/2,+1/2}  \\
( \sqrt{J}   f_{21} \; + \; \sqrt{J+1} f_{22} ) \; D^{J+1/2}_{-M-1/2,-1/2}  \\
( \sqrt{J+1} f_{41} \; + \; \sqrt{J}   f_{42} ) \;
D^{J+1/2}_{-M-1/2,+1/2}
\end{array} \right |   \; ,
$$
$$
(V^{(J+1/2)}_{i3} ) =   \sqrt{{J-M+1\over 2J+1}} \left |
\begin{array}{c}
( \sqrt{J}   f_{13} \; + \; \sqrt{J+1}  f_{14} ) \; D^{J+1/2}_{-M+1/2,-1/2}  \\
( \sqrt{J+1} f_{23} \; + \; \sqrt{J}    f_{24} ) \; D^{J+1/2}_{-M+1/2,+1/2}  \\
( \sqrt{J}   f_{33} \; + \; \sqrt{J+1}  f_{34} ) \; D^{J+1/2}_{-M+1/2,-1/2}  \\
( \sqrt{J+1} f_{43} \; + \; \sqrt{J}    f_{44} ) \;
D^{J+1/2}_{-M+1/2,+1/2}
\end{array}  \right |      \; ,
$$
$$
(V^{(J+1/2)}_{i4} )  = \sqrt{{J+M+1 \over 2J+1}} \left |
\begin{array}{c}
( \sqrt{J}   f_{13} \; + \; \sqrt{J+1} f_{14} ) \; D^{J+1/2}_{-M-1/2,-1/2}  \\
( \sqrt{J+1} f_{23} \; + \; \sqrt{J}   f_{24} ) \; D^{J+1/2}_{-M-1/2,+1/2}  \\
( \sqrt{J}   f_{33} \; + \; \sqrt{J+1} f_{34} ) \; D^{J+1/2}_{-M-1/2,-1/2}  \\
( \sqrt{J+1} f_{43} \; + \; \sqrt{J}   f_{44} ) \;
D^{J+1/2}_{-M-1/2,+1/2}
\end{array} \right |      \; .
$$

Now the functions  $f_{ab}$ in  (1.3)  should be  taken in accordance with the substitutions given below
$$
f_{ab} =   \left | \begin{array}{rrrr}
\lambda\; (K+iL)  & \lambda\; (M+iN)  &  \;\;\; (K+iL) & \;\;\;(M+iN) \\
\lambda\; (M-iN)  & \lambda\; (K-iL)  &  \;\;\; (M-iN) & \;\;\;(K-iL) \\
\Delta \; (K-iL)  & \Delta\; (M-iN)   &  \Delta \lambda\; (K-iL) &
\Delta \lambda  (M-iN) \\
\Delta \; (M+iN)  & \Delta \; (K+iL)   & \Delta \lambda (M+iN) &
\Delta  \lambda  (K+iL)
\end{array} \right |    \; ,
$$

\noindent where е $\lambda  = \pm 1$  and  $\Delta  = \pm 1$. Thus, from  (1.3) it follows
$$
V^{(J-1/2)}_{\epsilon JM\Delta \lambda } = \left |
\begin{array}{rrrr}
\lambda  \; \Omega    & - \lambda\; \Xi  &  \Omega   &  \Xi  \\
-\lambda \; \Upsilon  & \lambda \;  Z    & \Upsilon  &   Z   \\
\Omega   & \Xi  & \lambda \; \Omega &  - \lambda \;  \Xi  \\
\Upsilon & Z  &   - \lambda \; \Upsilon  & \lambda\;  Z
\end{array}  \right |   \left.  \begin{array}{l}
\leftarrow \;\;\; H^{+}  \\ \leftarrow \;\;\; H^{-} \\
\leftarrow \Delta  H^{+}  \\ \leftarrow \Delta  H^{-}
 \end{array} \right. \; ,
\eqno(2.4a)
$$

\noindent where symbols  $\Omega  ,\; \Xi  ,\; \Upsilon  ,\; Z$
stand for expressions
$$
\Omega = \sqrt{{J+M \over J(J+1)}} \; D^{J-1/2}_{-M+1/2,-1/2} \;\;
, \; \; Y = \sqrt{{J+M \over J(J+1)}}      \;
D^{J-1/2}_{-M+1/2,+1/2} \;\;  ,
$$
$$
\Xi  = \sqrt{{J-M \over J(J+1)}}   \; D^{J-1/2}_{-M-1/2,-1/2} \;\;
, \; \; Z    = \sqrt{{J-M \over J(J+1)}}   \;
D^{J-1/2}_{-M-1/2,-1/2} \;\;  ,
$$
$$
\eqno(2.4b)
$$

\noindent and  $H^{\pm }(r)$ represent
$$
H^{\pm }(r) = \sqrt{{J(J+1) \over 2J+1}} \; \left [\; \sqrt{J+1}\;
(K \pm  i L)\; - \; \sqrt{J}\; (M \pm  i N)\; \right  ]\;     .
\eqno(2.4c)
$$

In the same manner for  $V^{(J+1/2)}_{\epsilon JM}$  we get
$$
V ^{(J+1/2)}_{\epsilon JM\Delta \lambda }  = \left |
\begin{array}{rrrr}
\lambda \;  \Omega   &  \lambda\;  \Xi  &  \Omega    &   \Xi  \\
\lambda \; \Upsilon  &  \lambda\;   Z   &  \Upsilon  &    Z   \\
\Omega    &   \Xi    &   \lambda \;  \Omega    & \lambda \;  \Xi  \\
\Upsilon  &   Z       &  \lambda \;  \Upsilon  & \lambda \; Z
\end{array} \right |         \left. \begin{array}{l}
\leftarrow \;\;\; H^{+}  \\   \leftarrow \;\;\; H^{-}   \\
\leftarrow \Delta\;  H^{+} \\
\leftarrow \Delta\;  H^{-}    \end{array} \right. \; ,
\eqno(2.5a)
$$

\noindent where now symbols  $\Omega  ,\; \Xi  , \; \Upsilon
\;, Z$   note the following expressions  (compare with  $(2.4b)$):
$$
\Omega  = \sqrt{{J-M+1 \over J(J+1)}}\;
D^{J+1/2}_{-M+1/2,-1/2}\;\; , \;\; \Xi  = \sqrt{{J+M+1 \over
J(J+1)}}\; D^{J+1/2}_{-M-1/2,-1/2}\;\;  ,
$$
$$
\Upsilon = \sqrt{{J-M+1 \over J(J+1)}}\;
D_{-M+1/2,+1/2}^{J+1/2}\;\;  , \;\; Z = \sqrt{{J+M+1 \over
J(J+1)}} \; D^{J+1/2}_{-M-1/2,-1/2}\; ,
$$
$$
\eqno(2.5b)
$$

\noindent and
$$
H^{\pm}(r) = \sqrt{{J \over 2J+1}}\left [\; \sqrt{J+1}\; (K \pm  i
L)\; + \; \sqrt{J+1}\; (M \pm  i N)\; \right  ]\; . \eqno(2.5c)
$$

Now we should take into account substitution  $(1.8a,b)$ , which results in
$$
U^{I}_{\epsilon JM\Delta \lambda }  \; \rightarrow\; \{\;
V^{(J+1/2)}_{\epsilon JM\Delta \lambda } \; , \;\;
V^{(J-1/2)}_{\epsilon JM\Delta \lambda }  = 0\;  \}\; ,
$$

$$
U^{II}_{\epsilon JM\Delta \lambda } \; \rightarrow\;
 \{\;  V^{(J+1/2)}_{\epsilon JM\Delta \lambda }  = 0\; ,\;\;
 V^{(J-1/2)}_{\epsilon JM\Delta \lambda } \; \}  \; .
\eqno(2.6)
$$

Let us expand  rows $(4\times 4)$-matrices
 $V^{(J\pm 1/2)}_{\epsilon JM\Delta \lambda }(x)$  in terms of  Dirac solutions $\Psi _{\epsilon jm\delta }(x)$.
 First consider  $V^{(J+1/2)}(x)$.  We should take   $( \Delta ,\; \lambda )$
subsequently as
$$
 +1,  +1 ;\;\;   +1, -1; \;\;  -1, +1; \; \; -1, -1
$$

 \noindent and should  note in expressions  $(2.4a)$ and  $(2.4b)$ only signs at $\Omega
 ,\; \Xi ,\; Y ,\; Z$. Thus, we get
 $$
 V^{(J+1/2)}_{JM,+1,+1} =
\left | \begin{array}{cccc} +& +& +& + \\  + & + & + & + \\   + &
+ &
 + & + \\  + & + & + &  + \end{array} \right | \; \; ,  \;\;
V^{(J+1/2)}_{JM,+1,-1}  =   \left | \begin{array}{cccc} -& -& +& +
\\  - & - & + & + \\   + & + & - & - \\  + & + & - &  -
\end{array} \right | \;\; ,
$$
$$
V^{J+1/2}_{JM,-1,+1} =  \left | \begin{array}{cccc} +& +& +& + \\
+ & + & + & + \\   - & - & - & - \\  - & - & - &  -
\end{array} \right | \;\; ,  \;\;
V^{J+1/2}_{JM,-1,-1} =  \left | \begin{array}{cccc} -& -& +& + \\
- & - & + & + \\   - & - & + & + \\  - & - & + &  +
\end{array} \right | \; .
\eqno(2.7)
$$

 The functions $f$  and $g$ entering the matrix  $V^{(J+1/2)}_{\epsilon JM,+1,+1} (x)$ obey eqs.
  $(1.9a)$   (see also   $(1.10b)$).
Comparing equations for $f$  and $g$  with those for $F$ and  $G$ in (2.1),   and also noting relevant Wigner functions,
we conclude that
the row from the matrix  (2.7) satisfy the ordinary Dirac equation;
at this we find corresponding quantum number explicit form of linear expansions

$$
V^{(J+1/2)}_{JM,+1,+1}  =  \left \{ \; \sqrt{{J-M-1 \over
J(J+1)}}\; \left [\; \Psi ^{(1)}_{J+1/2,M-1/2,+1} \;+\; \Psi
^{(3)}_{J+1/2,M-1/2,+1}\; \right  ] \right.
$$
$$
\qquad \qquad  \left.  + \sqrt{{J+M+1 \over J(J+1)}}\; \left  [\; \Psi _{J+1/2,M+1/2,+1}
^{(2)} \;+\; \Psi ^{(4)}_{J+1/2,M+1/2,+1} \; \right ]\;  \right \}
\; . \eqno(2.8)
$$

\noindent In the same manner we consider three remaining cases.

Let us introduce the notation
$$
j = (J + 1/2)\; ,\; m = (M + 1/2)\;\; , \;\;  m' = M - 1/2\; ,
$$
$$
\alpha  = \sqrt{{J-M+1 \over J(J+1)}}\;\; , \;\; \beta  =
\sqrt{{J+M+1 \over J(J+1)}}\;\;  , \eqno(2.9a)
$$

\noindent them expansions of the Dirac-K\"{a}hler boson solutions in terns of fermion Dirac solutions
can be presented as follows
$$
V^{(J+1/2)}_{JM,+1,+1}  =  \alpha \; ( \Psi ^{(1)}_{jm',+1}\; + \;
\Psi ^{(3)}_{jm',+1} )\; + \; \beta\;  (\Psi ^{(2)}_{jm,+1}\; +\;
\Psi ^{(4)}_{jm,+1} )\; \;  ,
$$
$$
V^{(J+1/2)}_{JM,-1,-1}  =  \; \alpha \; (-\Psi ^{(1)}_{jm',+1}\; +
\; \Psi ^{(3)}_{jm',+1} )\; + \; \beta\;  (-\Psi ^{(2)}_{jm,+1} \;
+\;  \Psi ^{(4)}_{jm,+1} )\; \; ,
$$
$$
V^{(J+1/2)}_{JM,+1,-1}  =  \; \alpha \; (-\Psi ^{(1)}_{jm',-1}\; +
\; \Psi ^{(3)}_{jm',-1} )\; +\; \beta\;  (-\Psi ^{(2)}_{jm,-1}\;
+\;  \Psi ^{(4)}_{jm,-1} )\; \;  ,
$$
$$
V^{(J+1/2)}_{JM,-1,+1}  =  \; \alpha \; ( \Psi ^{(1)}_{jm',-1} \;
+\; \Psi ^{(3)}_{jm' ,-1} ) \; + \; \beta \;  ( \Psi
^{(2)}_{jm,-1}\; + \; \Psi ^{(4)}_{jm,-1} )\; \; . \eqno(2.9b)
$$

Not, let us consider solutions  $V^{(J-1/2)}_{JM\Delta \lambda }$ with the structure (only the first row is written down)
$$
V^{(J-1/2}_{JM,+1,+1} = \left | \begin{array}{rrrr}
(f + i g )\; \Omega  & . & . & . \\
- (f - i g ) \;\Upsilon   & . &  . &  .  \\
(f - i g ) \; \Omega  & . & . & . \\
- (f + i g ) \; \Upsilon  & . & . & .     \end{array} \right |  \;
; \eqno(2.10a)
$$

\noindent where  $f$ and $g$ obey
$$
( {d \over dr}  - {J \over  r} ) f + (\epsilon + m) g  = 0   \;\;
,
$$
$$
({d  \over dr}  + {J \over r}  ) g - (\epsilon - m) f  = 0 \; .
\eqno(2.10b)
$$

\noindent Note that eqs. $(2.10b)$  do not coincide with relevant equation in Dirac case
at  $\delta  = \pm  1$ ; besides, in the rows of the matrix  $V^{(J-1/2)}_{JM,+1,+1}$ (in $(2.10b)$
on;y the first row is written down) we do not have the structure  required  to relate them with Dirac
solutions
$$
\Psi _{jm,\delta = +1} \sim   \left | \begin{array}{c} + \\ +  \\ +
\\ +   \end{array} \right | \; , \;\; \Psi _{jm,\delta = -1} \sim
\left | \begin{array}{c}
 + \\ +  \\ -  \\ -    \end{array} \right | \; .
$$

\noindent However, both impediments can be removed by simple change in notation
  $f = - G , g = + F$. Then
$$
(f + i g) = i (F + i G)\, \qquad
(f - i g) = - i (F - i G )\;,
$$
\noindent and eqs.  $(2.10b)$ read as
$$
({d \over dr} + {J \over r}) F + ( \epsilon  - m ) G = 0 \;\; ,
$$
$$
({d \over dr} - {J \over r}) G - ( \epsilon  + m ) F = 0 \;\; ;
\eqno(2.11a)
$$

\noindent and $(2.10a)$ reduces to
$$
V^{(J-1/2)}_{JM,+1,+1}  =     \left | \begin{array}{cccc}
i (F + i G )\; \Omega  & . & . & . \\
i (F - i G ) ...\Upsilon   & . &  . &  . \\
- i (F - i G )\;  \Omega   & . &  .  & .  \\
- i (F + i G )\; \Upsilon  & . & . & .     \end{array} \right | \;
. \eqno(2.11b)
$$

\noindent Further, allowing for  (2.11a,b) and structure of the matrix in four cases
$$
V^{(J-1/2)}_{JM,+1,+1}    = \left | \begin{array}{cccc}
+ &  - &   + &  -   \\
+ &  - &   + &  -   \\
- &  + &   - &  +   \\
- &  + &   - &  +      \end{array} \right |   \; , \;\;\;
V^{(J-1/2}_{JM,+1,-1} = \left | \begin{array}{cccc}
- &  + &   + &  -   \\
- &  + &   + &  -   \\
- &  + &   + &  -   \\
- &  + &   + &  -      \end{array} \right | \; ,
$$
$$
V^{(J-1/2}_{JM,-1,+1} = \left | \begin{array}{cccc}
+ &  - &   + &  -   \\
+ &  - &   + &  -   \\
+ &  - &   + &  -   \\
+ &  - &   + &  -      \end{array} \right |  \; , \;\;\;
V^{(J-1/2)}_{JM,-1,-1} = \left | \begin{array}{cccc}
- &  + &   + &  -   \\
- &  + &   + &  -   \\
+ &  - &   - &  +   \\
+ &  - &   - &  +      \end{array} \right | \; ,
$$
$$
\eqno(2.12a)
$$

\noindent we arrive at needed expansions
for $V^{(J-1/2)}_{JM\Delta \lambda }$
$$
V^{(J-1/2)}_{JM,+1,-1}  =   \; \rho \; ( i \Psi ^{(1)} _{jm' ,-1}
\; + \; i\Psi ^{(3)}_{jm',-1} ) \; -  \; \sigma \; (i \Psi
^{(2)}_{jm,-1}\; +\; i \Psi ^{(4)}_{jm,-1} )\;   ,
$$
$$
V^{(J-1/2)}_{JM,-1,-1}   =   \; \rho \; (-i\Psi ^{(1)}_{jm',-1}\;
+\;
 i\Psi ^{(3)}_{jm',-1} )\;  + \;
\sigma \;  (i \Psi ^{(2)}_{jm,-1}\;  -\; i \Psi ^{(4)}_{jm,-1} )\;
,
$$
$$
V^{(J-1/2}_{JM,+1,-1}  =  \;
 \rho \; (-i \Psi ^{(1)}_{jm',+1}\; + \;
i \Psi ^{(3)}_{jm',+1} ) \;  +  \; \sigma \; (i\Psi ^{(2)}_{jm,+1}
\; - \; i \Psi ^{(4)}_{jm,+1} )\;  ,
$$
$$
V^{(J-1/2)}_{JM,-1,+1}  =   \; \rho \; (i \Psi ^{(1)}_{jm',+1} \;
+ \; i \Psi ^{(3)}_{jm',+1} ) \;  - \; \sigma \;  (i \Psi
^{(2)}_{jm,+1} \; + \;  i \Psi ^{(4)}_{jm,+1} )
 \; ,
$$
$$
\eqno(2.12b)
$$

\noindent where
$$
j = (J-1/2)\;\; , \;\; \rho  = \sqrt{{J+M \over J(J+1)}}\;\; ,\;\;
\sigma  = \sqrt{{J-M \over J(J+1)}} \; .
$$

 Expansions for the case of minimal value $J=0$
 will be much more simple
 $$
V_{00,+1,+1}  =  \;
 \Psi ^{(1)}_{1/2,-1/2,+1}\; +\;
\Psi ^{(3)} _{1/2,-1/2,+1} \; + \;
 \Psi ^{(2)}_{1/2,+1/2,+1} \;+\; \Psi ^{(4)}_{1/2,+1/2,+1} \; ,
  $$
 $$
V_{00,-1,-1}  =   \; - \Psi ^{(1)}_{1/2,-1/2,+1}\; +\;
 \Psi ^{(3)}_{1/2,-1/2,+1} \; + \;
- \Psi ^{(2)}_{1/2,+1/2,+1}\; +\; \Psi ^{(4)}_{1/2,+1/2,+1} \;
,
$$
$$
V_{00,+1,+1}   =    \; - \Psi ^{(1)}_{1/2,-1/2,-1}\; +\;
 \Psi ^{(3)}_{1/2,-1/2,-1}\; + \;
 - \Psi ^{(2)}_{1/2,+1/2,-1} \; + \; \Psi ^{(4)}_{1/2,+1/2,-1}\;   ,
 $$
 $$
 V_{00,+1,+1}   =   \;
 \Psi ^{(1)}_{1/2,-1/2,-1} \; +\;
\Psi ^{(3)}_{1/2,-1/2,-1} \; +  \;
 \Psi ^{(2)}_{1/2,+1/2,-1} \; + \; \Psi ^{(4)}_{1/2,+1/2,-1} \;  .
\eqno(2.13)
$$

\section{Discussion}

Above, after performing a special transformation over $4 \times 4$-matrix $U(x) \Longrightarrow V(x)$, spherical boson
solution of the Dirac--K\"{a}hler equation, simple linear expansions of the four rows of new representative
of the Dirac--K\"{a}hler field $V(x)$ in terms of spherical fermion solutions $\Psi_{i}(x)$  of the four ordinary Dirac equation
have been derived. However, this fact cannot be interpreted as the  possibility not to  distinguish between
the Dirac-K\"{a}hler fileld and the system four Dirac fermions.
The main formal argument  is that the special transformation $(S(x) \otimes I)$ involved  does not belong
to the group of tetrad local gauge transformation for Dirac-K\"{a}hler field, 2-rank bispinor under the Lorentz group.
Therefore, the linear expansions between boson and fermion functions are not gauge invariant under the
group of local tetrad rotations.

Formal possibility to produce such expansions exists  only for the case of flat Minkowski space-time,
and cannot be extended to  any  other space-time with curvature.
For instance, let us specify the situation  for spherical space with constant positive curvature
In spherical coordinates and tetrad

$$
dS^{2}  =  dt^{2}  - d\chi ^{2} - \sin  ^{2}\chi  (\; d\theta ^{2}
+ \sin ^{2} \theta  d \phi ^{2}\; ) \;  \;  ,
$$
$$
e^{\alpha }_{(0)} = (1, \; 0 ,\;  0 ,\;  0 )\; , \;\; e^{\alpha
}_{(1)} = ( 0 ,\; 0 ,\; \sin^{-1}\chi  ,\;  0 )\;  ,
$$
$$
e^{\alpha }_{(2)} = (0,\; 0, \; 0, \;  \sin^{-1} \chi \sin^{-1}
\theta  ) \; ,  \; e^{\alpha }_{(3)} = ( 0 , 1 , 0 , 0 ) \; ,
\eqno(3.1)
$$

\noindent the Dirac--K\"{a}hler
equation takes the form
$$
\left [\; i \gamma ^{0} \; \partial_{t}\; + \; i \left  (\gamma
^{3}\;
\partial_{\chi} \;  + \;
 {\gamma ^{1} J^{31} + \gamma ^{2} J^{32} \over \mbox{tg}\;   \chi  } \right )\;
+ \;{1 \over \sin  \chi } \; \Sigma _{\theta ,\phi }  -  m  \right
] \; U(x) =  0  \; . \eqno(3.2)
$$

Most of  calculations performed above are valid here with small formal changes;
in particular, instead of  $(1.7b)$ we have at
 $ \Delta =+1 \; , \;\; \lambda =+1  $
$$
{dK \over d\chi}  + {a \over \sin  \chi}  M + (\epsilon  + m) L  =
0 \; ,
$$
$$
{dL \over d\chi} - {a \over \sin \chi}  N - ( \epsilon - m) K  = 0
\; ,
$$
$$
({d \over d\chi} + {1 \over \mbox{tg}\;   \chi }) M + {a \over
\sin \chi} K + (\epsilon  + m) N  = 0 \; ,
$$
$$
({d \over d\chi} - {1 \over \mbox{tg}\; \chi} ) N  - {a \over \sin
\chi } L  - (\epsilon  - m) M  = 0 \; . \eqno(3.3)
$$

\noindent However, the above used substitutions of the form  $(1.8a,b)$,
cannon be imposed because they are not consistent with eqs. (3.3.
This means that no solutions for radial functions  of (formally)  fermion type  exist in spherical space.
The latter is due to the fact that one cannot find any transformation like $I \otimes S(x)$
which would divide  the Dirac--K\"{a}hler equation into four separate  Dirac equations.

\end{document}